\documentclass[conference]{IEEEtran}
\IEEEoverridecommandlockouts
\usepackage{cite}
\usepackage{amsmath,amssymb,amsfonts}

\usepackage{graphicx}
\usepackage{textcomp}
\usepackage{xcolor}
\usepackage{algorithm}
\usepackage{algpseudocode}

\usepackage{listings}
\usepackage{xcolor}
\def\BibTeX{{\rm B\kern-.05em{\sc i\kern-.025em b}\kern-.08em
    T\kern-.1667em\lower.7ex\hbox{E}\kern-.125emX}}
\begin{document}

\title{SFDS: Selective File Disclosure System}

\author{\IEEEauthorblockN{Aditya Mitra}
\IEEEauthorblockA{\textit{CyberMACS} \\
\textit{Kadir Has University}\\
Istanbul, Turkey \\
aditya.mitra@stu.khas.edu.tr \\
ORCiD: 0000-0002-9612-0810}
\and
\IEEEauthorblockN{Quazi Fariha Tasnim}
\IEEEauthorblockA{\textit{CyberMACS} \\
\textit{Kadir Has University}\\
Istanbul, Turkey \\
qtasnim@stu.khas.edu.tr}
\and
\IEEEauthorblockN{Hristina Mihajloska Trpcheska}
\IEEEauthorblockA{\textit{Faculty of Computer Science and Engineering} \\
\textit{Ss. Cyril and Methodius University}\\
Skopje, North Macedonia \\
hristina.mihajloska@finki.ukim.mk}
}

\maketitle

\begin{abstract}
Access control to networked resources has been a longstanding challenge. The conventional solution relies on authentication mechanisms, which introduce additional complexities associated with Identity and Access Management (IAM). Such systems require user authentication, identity management, and authorization services, while also introducing security risks arising from vulnerabilities, misconfigurations, or implementation flaws. Furthermore, different file formats employ different mechanisms for ensuring authenticity and integrity through digital signatures. For example, PDF documents support the PDF Advanced Electronic Signature (PAdES) standard, whereas plain text files typically lack a standardized mechanism for embedding digital signatures. This paper proposes an architecture based on the Selective Disclosure JSON Web Token (SD-JWT) standard for securely sharing read-only files. The proposed architecture embeds cryptographic signatures and integrity protection directly into the shared resource, providing verifiable authenticity without relying on complex IAM infrastructures, such as centralized user databases, authentication services, or authorization mechanisms. By eliminating these components, the proposed solution simplifies deployment while maintaining strong security guarantees for the distribution of immutable resources.
\end{abstract}

\begin{IEEEkeywords}
privacy, decentralize, JWT, SD-JWT, Verifiable credentials
\end{IEEEkeywords}

\section{Introduction}
In many scenarios read-only files are shared but an extensive access control system is used for the process. This involves the use of databases, authentication systems, and authorization lists, followed by extensively securing the authentication systems. But even, once the files are downloaded by the user, proving the integrity of the file and the source requires a whole separate infrastructure for digital signatures. 

This scenario is often seen in educational institutions, where they often need to maintain authentication systems for the student grade sheets, ensuring one student cannot access that of another. However, the transcripts still need to be digitally signed, requiring one more digital signature infrastructure and signing the transcript of each student individually.

Another appropriate example would be the medical industry where maintaining the integrity and confidentiality of patient data is very important. However, access control of such Personal Health Information (PHI) often involves creating robust authentication and authorization systems \cite{Quazi2024-ot}. Further, maintaining the integrity and authenticity of such data is paramount for proper treatment of the patients. However, most medical reports are generated using legacy medical machines which do not cryptographic digital signatures and integrity checking mechanisms. Further, the file formats used by such legacy machines do not support such digital signature or other methods of maintaining integrity and authenticity.

This study proposes a system to simplify these problems, leveraging the SD-JWT standard \cite{rfc9901} for sharing such information. It ensures only the authorized holders holding the Verifiable Presentation corresponding to the information may only be able to access the contents of the information, and at the same time provide authenticity and integrity from the issuer.

The contributions in this study include:
\begin{itemize}
    \item Development of an architecture to share files over SD-JWT and Verifiable Credentials standard.
    \item Ensure interoperability with SD-JWT and Verifiable Presentations.
    \item Ensure the Verifiable Presentations required to access the files are small enough to be transferred over limited bandwidth or stored on devices with limited storage or other formats like QR Codes. 
    \item Ensure the integrity of the files shared this way is maintained and is verifiable.
\end{itemize}

\section{Literature Review}
Access control and digital signature to big files irrespective of their types is important in modern internet. While modern file types like PDFs have signature standards like PAdES \cite{pades}, legacy file formats often lack digital signature and integrity mechanisms. There have been multiple studies highlighting problems regarding privacy and integrity of legacy standards like text, images, etc. A study \cite{fase} highlights fine-grained access control of multimedia content with caching in network. Another study \cite{bigdata} highlights the challenges on Big Data access control schemes.

On the other hand, the SD-JWT standard \cite{rfc9901} along with W3C Verifiable Credentials \cite{w3cvc} gives a promising standard for securely disclosing claims which are verifiable cryptographically. Claims can be represented as read only values, usually a string or number and often represents the credential of the user. For example, identity cards, driving license, student records, etc., can all be represented in the form of verifiable credentials model. Most ID documents are, in fact, stored and shared as verifiable credentials as per the eIDAS (electronic Identification, Authentication and Trust Services) regulation under the European Union \cite{eidassurvey} \cite{eidaswallet}. 
The SD-JWT standard has been instrumental in maintaining the privacy of Verifiable Credentials. It allows a presenter to reveal only a part of the verifiable credential with he needs to prove to the verifier while keeping the rest of it a secret. It has been instrumental in preserving privacy of the credentials while at the same time being able to use only the ones required. \cite{sdjwtssi} \cite{decenid}.

One of the major problems with the verifiable credential model, is that the claims used in the same are almost always only string or numbers. This brings in a bigger hurdle when it is needed to share big files, usually binary files in a verifiable manner. One of the easier solutions could have been encoding the files to Base64 and use that as a string value, analogous to standard Verifiable credentials. However, it would eventually bloat up the presentations and make it difficult to share them over media that offers very little space, like QR codes or smart cards. One of the main ideas of SD-JWT based credentials, accompanying physical cards is being presentable by the card only, and it often boils down to QR codes and NFC only. \cite{blockchainid} \cite{novidchain}.

On the other hand, often authentication and access management are involved with privately sharing files. This involves authentication by standard factors like knowledge based, possession based or inherence based \cite{sp80063b}. This requires complex architecture involving databases, authentication systems, and so on. The files are often shared on cloud, IPFS etc., without a standard for selective disclosure. \cite{turtl} \cite{smartcity}. The proposed Selective File Disclosure System outlines a procedure to privately share files in a private manner where the content of the file remains verifiable. Table \ref{tab:sfds_comparison} highlights the advantages of SFDS over traditional methods.

\begin{table}[htbp]
\caption{Comparison of SFDS with Existing Approaches}
\begin{center}
\renewcommand{\arraystretch}{1.3}
\begin{tabular}{|p{1.6cm}|p{0.6cm}|p{1.4cm}|p{1.6cm}|p{1.6cm}|}
\hline
\textbf{Feature} & \textbf{SFDS} & \textbf{File Sharing with Standard IAM} & \textbf{SD-JWT} & \textbf{Use of Blockchain or distributed file system like IPFS} \\
\hline

Big file sharing capability &
Yes &
Yes &
No &
Yes \\
\hline

Privacy preserving architecture &
Yes &
Yes &
Yes &
Partial \\
\hline

Simplicity of architecture &
Yes &
No &
Yes &
No \\
\hline

Requires complex moving parts like databases &
No &
Yes &
No &
No \\
\hline

Revocability &
Yes &
Yes &
Partial: revoked records only hamper the verifiability, not the content &
Partial: depends on implementation \\
\hline

Limited storage: Can big files be presented on limited storage media &
Yes &
Yes (or no storage media required if memorized secrets are used for authentication) &
No (if files are shared in presentations) &
Partial: depends on implementation \\
\hline

\end{tabular}
\label{tab:sfds_comparison}
\end{center}
\end{table}

\section{Methodology}
\begin{figure*}[htbp]
\centering
\includegraphics[width=\textwidth]{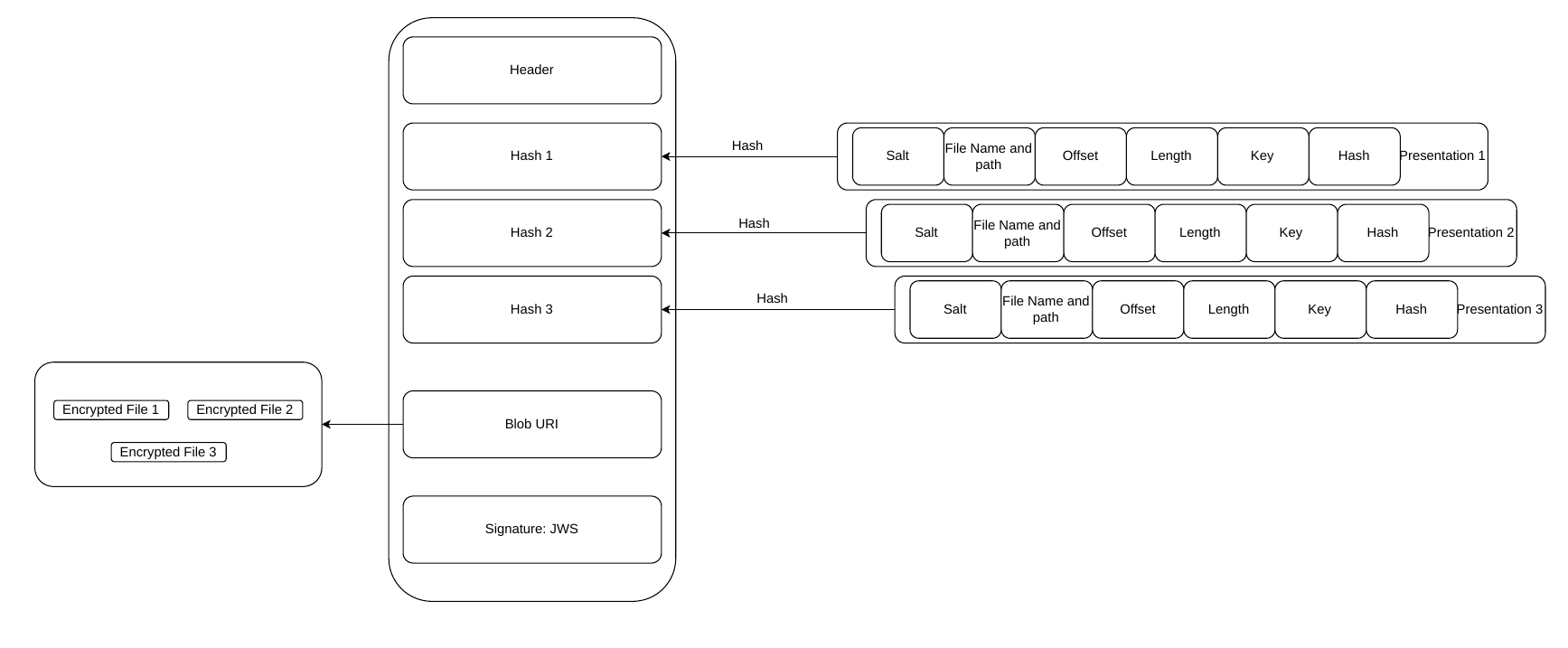}
\caption{SFDS Structure}
\label{fig:structure}
\end{figure*}

The proposed architecture involves three actors: the issuer, the presenter and the verifier. The issuer is the source of the data and ensures the authenticity of the data. On the other hand, the presenter or the holder is the entity that receives the data from the issuer and may hold it for future use. The verifier is the entity requiring proof of the data. This follows the W3C Verifiable Credential Data Model, ensuring all claims are verifiable cryptographically. This proposed system expands on the Verifiable Credential and SD-JWT models to ensure the same can be expanded to bigger files, instead of smaller claims that can fit in one single JSON file. It further streamlines the issuance for multiple files together. For example, an educational institute may create one JSON file containing the transcripts of all students and share the respective presentations to the students who may use it to read their transcripts and at the same time share it with potential verifiers like recruiters, who may verify the validity and read the transcript. The system also ensures these transcripts can be issued in human-readable formats like PDF or scanned copies like JPEG instead of only machine-readable JSON. This ensures the shared information may retain non-text information as well, for example scanned records of student answer scripts. Figure \ref{fig:structure} demonstrates the structure of a SFDS record.

\subsection{Issuer flow}
The issuer is the entity that produces the files and is responsible for maintaining authenticity. The issuer creates the JSON Web Token containing the information of the files, as well as the presentation that grants the holder permission to view the files and verify its authenticity.

The issuer iterates through the directory containing the files to share. A cryptographic hash of the file is generated. It generates a random symmetric cryptographic key for each file and uses AES in Galois Counter Mode with 256-bit key size to encrypt each file separately. While the current implementation of the system uses AES-GCM, other implementations may use other symmetric cryptographic algorithms as well that provide strong security, based on the speed and performance requirements. The 'Alg' field in the JWT identifies the used algorithm by its COSE identifier. AES-GCM provides confidentiality and ciphertext integrity through its AEAD authentication tag. The plaintext hash included in the SD-JWT disclosure provides an additional issuer-authenticated content binding and enables verification that the decrypted file corresponds to the issuer's intended file. Therefore, the AEAD tag and the plaintext hash serve complementary purposes and both must be verified. The encrypted file is appended to a binary large object (Blob). The offset at which the encrypted file is appended, and the length of the encrypted file are recorded. The disclosure is made which contains the file name, the offset, the length of the encrypted file, the encryption key, hash of the original file and a random salt.

These disclosures are hashed and added to the JWT. The Blob created this way may contain random padding between two encrypted files, before all files or at the end to reduce chances of inference-based attacks. The blob could be hosted on a web platform or a decentralized platform and the URI to access it is added to the JWT. The JWT is then signed by the issuer key to create the JWS which is appended to the end.
The issuer may then share the JWT and the respective disclosures with the authorized holders. The flow of the issuer may be represented in algorithm \ref{alg:sfds_issuer}.

\begin{algorithm}
\caption{Issuer Workflow}
\label{alg:sfds_issuer}
\begin{algorithmic}
\State $(Issuer_{pub}, Issuer_{priv}) \gets \textsc{KeyGen}(\text{ECDSA})$
\State $Blob \gets \emptyset$
\State $SDList \gets \emptyset$
\State $Disclosures \gets \emptyset$

\ForAll{files in the directory and its subdirectories}

    \State $(FileName, File) \gets \textsc{ReadFile}()$
    \State $Hash \gets \textsc{SHA256}(File)$
    \State $Key \gets \textsc{RandomKey}()$
    \State $Nonce \gets \textsc{RandomNonce}()$
    \State $(Cipher, Tag) \gets \textsc{AES\mbox{-}GCM\mbox{-}256\_Encrypt}
    (File, Key, Nonce)$
    \State $EncryptedFile \gets Cipher \parallel Nonce \parallel Tag$

    \State $Offset \gets \textsc{Length}(Blob)$
    \State $Length \gets \textsc{Length}(EncryptedFile)$
    \State $Alg \gets 3$ \Comment{COSE identifier for AES-GCM-256}

    \State $FileData \gets \{FileName, Key, Hash, Offset, Length,
    Alg\}$

    \State $Blob \gets Blob \parallel EncryptedFile$

    \State $Salt \gets \textsc{RandomSalt}()$

    \State $SDList \gets SDList \cup
           \{\textsc{SHA256}(Salt \parallel FileData)\}$

    \State $Disclosures \gets Disclosures \cup
           \{Salt \parallel FileData\}$

\EndFor

\State $BlobURI \gets \textsc{URI}(Blob)$

\State $Content \gets Header \parallel SDList \parallel BlobURI$

\State $JWT \gets Content \parallel
       \textsc{Sign}(Content, Issuer_{priv})$

\State \Return $(Issuer_{pub}, JWT, Disclosures)$
\end{algorithmic}
\end{algorithm}

When the issuer has to revoke a file, apart from standard revocability procedures outlined in W3C Verifiable Credentials model, the issuer may replace the corresponding bytes in the blob with random data to ensure it cannot be retrieved and decrypted, without breaking the integrity of other files in the blob.

\subsection{Presenter flow}

The presenter is the authorized holder of the disclosures and may present it to verifiers. If the presenter attempts to view the content of the file referred to in the disclosure, the presenter will follow the verifier flow. The presenter may decide which files to disclose to the verifier, followed by creating a verifiable presentation in accordance with the W3C VC Model. The presenter flow may be represented in Algorithm \ref{alg:sfds_presenter}.

\begin{algorithm}
\caption{Presenter Workflow}
\label{alg:sfds_presenter}
\begin{algorithmic}
\State $Disclosures \gets$ disclosures received from issuer
\State $PresentationList \gets \emptyset$

\ForAll{$Disclosure \in Disclosures$}

    \State $FileData \gets \textsc{ExtractFileData}(Disclosure)$
    \State $FileName \gets \textsc{ExtractFileName}(FileData)$
    \State Prompt user whether to disclose $FileName$
    \If{user chooses to disclose}
        \State $PresentationList \gets PresentationList \cup \{Disclosure\}$
    \EndIf

\EndFor

\State \Return $PresentationList$
\end{algorithmic}
\end{algorithm}

\subsubsection{Verifier flow}
The verifier is the entity that holds the presentation and is able to read the content of the files referred to by them. The verifier validates the verifiable presentation, followed by extracting the presentation list from it. The verifier also validates the JWT with the public key of the issuer. This is followed by the verifier checking whether the hashes of the disclosures in the presentation are in the JWT to ensure the disclosures are valid.

This is followed by the verifier extracting the key, hash, offset and length from the disclosures. The verifier then fetches the part of the blob from the offset to the desired length. If it is shared on web, the Range header \cite{rfc7233} \cite{rfc9110} may be used to get only the desired part of the blob.

The part of the blob is then decrypted using the key. A hash of the decrypted blob is verified against the hash in the disclosure to ensure it is the intended file. This is followed by writing the file to the verifier disk as per the file name in the disclosure. The verifier may then be able to read the file and be assured of the integrity and authenticity of the file. The flow of the verifier might be represented by Algorithm \ref{alg:sfds_verifier}.

\begin{algorithm}
\caption{Verifier Workflow}
\label{alg:sfds_verifier}
\begin{algorithmic}
\State $PresentationList \gets$ received from presenter
\State $(Issuer_{pub}, JWT) \gets$ received from issuer

\State $(Content, Signature) \gets \textsc{ExtractJWT}(JWT)$

\State \textsc{VerifySignature}$(Signature, Content, Issuer_{pub})$

\State $(SDList, BlobURI) \gets \textsc{ExtractContent}(Content)$

\ForAll{$Presentation \in PresentationList$}

    \State Verify $\textsc{SHA256}(Presentation) \in SDList$

    \State $FileData \gets \textsc{ExtractFileData}(Presentation)$

    \State $(FileName, Key, Hash,$
    \Statex \hspace{\algorithmicindent}$Offset, Length, Alg)
           \gets \textsc{Parse}(FileData)$

    \State $EncryptedFile \gets$
    \Statex \hspace{\algorithmicindent}
    $\textsc{Fetch}(BlobURI, Offset, Offset + Length)$

    \State $Cipher, Nonce, Tag \gets \textsc{Unpack}(EncryptedFile)$

    \State $File \gets$
    \Statex \hspace{\algorithmicindent}
    $\textsc{AES\mbox{-}GCM\mbox{-}256\_Decrypt}(Cipher, Key, Nonce, Tag)$

    \State Verify $\textsc{SHA256}(File) = Hash$

    \State \textsc{WriteFile}$(FileName, File)$

\EndFor
\end{algorithmic}
\end{algorithm}

\section{Threat model}

The threat model of the proposed SFDS architecture encompasses the threat model of W3C Verifiable Credential data model. The contribution of this study adds sharing big binary files and Binary large objects over the verifiable credential model. This expands it to support non-text data which includes items like raw medical records produced by medical equipment.

The blob storage and share are an added dependency to SFDS compared to W3C Verifiable Credentials. The threat model of the blob storage share includes:
\begin{itemize}
    \item \textbf{Unauthorized access.} The blob storage is open to access without permission. Unauthorized access to the same reveals nothing because all data is encrypted.
    \item \textbf{Inference attacks.} Malicious actors may try to infer information by studying metadata like the blob storage size. To mitigate the same issuer may add random padding between two encrypted files or before and after files. This padding, being random, should be statistically not differentiable from the encrypted data, thus mitigating inference-based attacks. Further, the use of Decoy Disclosures in accordance with \cite{rfc9901} ensures inference attacks from the number of hashes present in the JWT is mitigated.
\end{itemize}

The security goals of the shared files are maintained by:

\begin{itemize}
    \item \textbf{Confidentiality.} The disclosure is required to decrypt the file. Without the disclosure, the files remain confidential.
    \item \textbf{Integrity.} The JWT is signed with the private key of the issuer. The validity of the signature ensures the integrity of the JWT. The integrity of the presentations is validated by comparing the cryptographic hash of the same with that on the JWT. The integrity of the files shared over SFDS is validated by comparing the hash of the file against the one in the presentations. Thus, integrity is established in a transitive way.
    \item \textbf{Availability.} Even if the Blob URI faces attacks related to availability, the presenter may separately supply file to the verifier, and the verifier would still be able to validate the authenticity and integrity of the file.
\end{itemize}

\section{Experimental setup}

The proposed system has been implemented in python and tested. Three independent computers were used as issuer, holder and verifier respectively. The issuer’s public key was available to the verifier, and the presenter supplied the presentation. In the experimental scenario, the issuer shared a folder with multiple multimedia, text and executable files. The presenter chose the files to be disclosed and created the presentation. The Verifier was able to recreate the shared files and verify the signature to validate the integrity. The issuer was then able to use the shared files. 

\begin{figure}[htbp]
\centering
\includegraphics[width=\columnwidth]{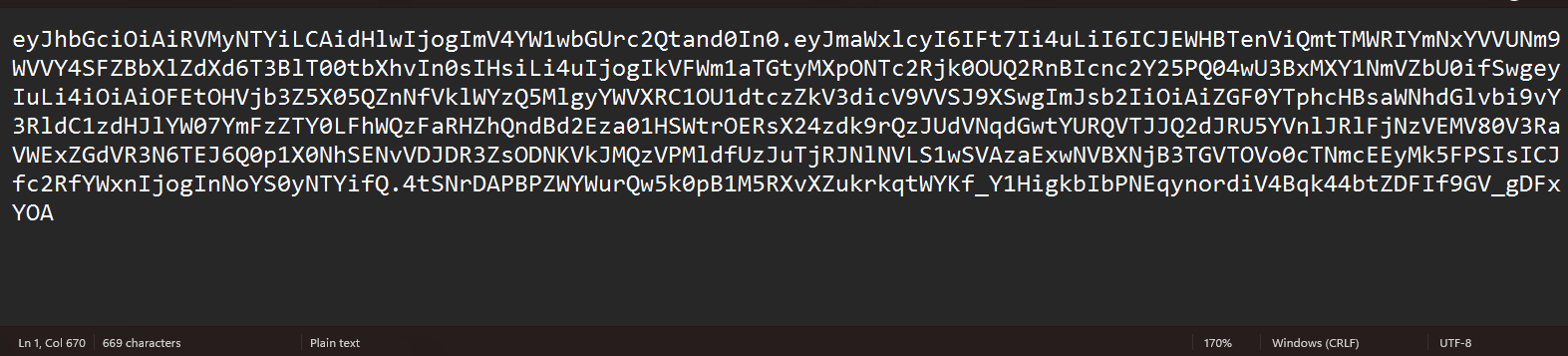}
\caption{JWT generated by implementation}
\label{fig:jwt}
\end{figure}

\begin{figure}[htbp]
\centering
\includegraphics[width=\columnwidth]{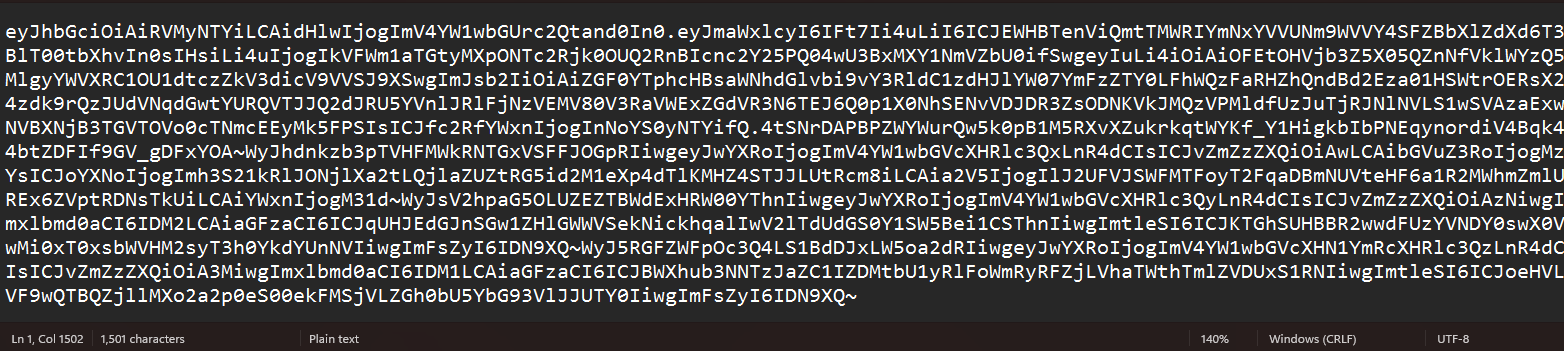}
\caption{Disclosures}
\label{fig:disclosure}
\end{figure}

This study does not include a performance study of the proposed standard since it is highly dependent on the implementation. Figure \ref{fig:jwt} shows the JWT generated by the implementation. The token contains disclosure digest placeholders under the files array and a blob reference used to retrieve the encrypted bundle. Figure \ref{fig:disclosure} shows the SD-JWT presentation produced by the holder, consisting of the issuer-signed JWT followed by selected Base64 URL-encoded disclosures separated by ASCII tilde characters.

The Python-based implementation and the test vectors for this study is available at \cite{gitrepo}.

\section{RFC 9901 compliance and Verifier Reconstruction}

This section highlights the test vectors and outputs from the sample implementation.

\begin{lstlisting}[caption={Issuer signed JWT in compact serialization}, label={lst:signedjwtpayload}]
eyJhbGciOiAiRVMyNTYiLCAidHlwIjogImV4YW1wbGUrc2Qtand0In0.eyJmaWxlcyI6IFt7Ii4uLiI6ICJEWHBTenViQmtTMWRIYmNxYVVUNm9WVVY4SFZBbXlZdXd6T3BlT00tbXhvIn0sIHsiLi4uIjogIkVFWm1aTGtyMXpONTc2Rjk0OUQ2RnBIcnc2Y25PQ04wU3BxMXY1NmVZbU0ifSwgeyIuLi4iOiAiOFEtOHVjb3Z5X05QZnNfVklWYzQ5MlgyYWVXRC1OU1dtczZkV3dicV9VVSJ9XSwgImJsb2IiOiAiZGF0YTphcHBsaWNhdGlvbi9vY3RldC1zdHJlYW07YmFzZTY0LFhWQzFaRHZhQndBd2Eza01HSWtrOERsX24zdk9rQzJUdVNqdGwtYURQVTJJQ2dJRU5YVnlJRlFjNzVEMV80V3RaVWExZGdVR3N6TEJ6Q0p1X0NhSENvVDJDR3ZsODNKVkJMQzVPMldfUzJuTjRJNlNVLS1wSVAzaExwNVBXNjB3TGVTOVo0cTNmcEEyMk5FPSIsICJfc2RfYWxnIjogInNoYS0yNTYifQ.4tSNrDAPBPZWYWurQw5k0pB1M5RXvXZukrkqtWYKf_Y1HigkbIbPNEqynordiV4Bqk44btZDFIf9GV_gDFxYOA
\end{lstlisting}

\begin{lstlisting}[caption={Decoded Issuer-signed JWT Payload}, label={lst:jwtpayload}]
{
	  "files": [
	    {
		      "...": "DXpSzubBkS1dHbcqaUT6oVUV8HVAmyYuwzOpeOM-mxo"
	    },
	    {
		      "...": "EEZmZLkr1zN576F949D6FpHrw6cnOCN0Spq1v56eYmM"
	    },
	    {
    	      "...": "8Q-8ucovy_NPfs_VIVc492X2aeWD-NSWms6dWwbq_UU"
	    }
	  ],
	  "blob": "data:application/octet-stream;base64,XVC1ZDvaBwAwa3kMGIkk8Dl_n3vOkC2TuSjtl-aDPU2ICgIENXVyIFQc75D1_4WtZUa1dgUGszLBzCJu_CaHCoT2CGvl83JVBLC5O2W_S2nN4I6SU--pIP3hLp5PW60wLeS9Z4q3fpA22NE=",
	  "_sd_alg": "sha-256"
}

\end{lstlisting}

\begin{lstlisting}[caption={SD-JWT presentation containing selected disclosures}, label={lst:disclosurelist}]
~WyJhdnkzb3pTVHFMWkRNTGxVSFFJOGpRIiwgeyJwYXRoIjogImV4YW1wbGVcXHRlc3QxLnR4dCIsICJvZmZzZXQiOiAwLCAibGVuZ3RoIjogMzYsICJoYXNoIjogImh3S21kRlJONjlXa2tLQjlaZUZtRG5id2M1eXp4dTlKMHZ4STJJLUtRcm8iLCAia2V5IjogIlJ2UFVJSWFMTFoyT2FqaDBmNUVteHF6a1R2MWhmZmlUREx6ZVptRDNsTkUiLCAiYWxnIjogM31d~WyJsV2hpaG5OLUZEZTBWdExHRW00YThnIiwgeyJwYXRoIjogImV4YW1wbGVcXHRlc3QyLnR4dCIsICJvZmZzZXQiOiAzNiwgImxlbmd0aCI6IDM2LCAiaGFzaCI6ICJqUHJEdGJnSGw1ZHlGWWVSekNickhqalIwV2lTdUdGS0Y1SW5Bei1CSThnIiwgImtleSI6ICJKTGhSUHBBR2wwdFUzYVNDY0swX0VwMi0xT0xsbWVHM2syT3h0YkdYUnNVIiwgImFsZyI6IDN9XQ~WyJ5RGFZWFpOc3Q4LS1BdDJxLW5oa2dRIiwgeyJwYXRoIjogImV4YW1wbGVcXHN1YmRcXHRlc3QzLnR4dCIsICJvZmZzZXQiOiA3MiwgImxlbmd0aCI6IDM1LCAiaGFzaCI6ICJBWXhub3NNTzJaZC1IZDMtbU1yRlFoWmRyRFZjLVhaTWthTmlZVDUxS1RNIiwgImtleSI6ICJoeHVLVF9wQTBQZjllMXo2a2p0eS00ekFMSjVLZGh0bU5YbG93VlJJUTY0IiwgImFsZyI6IDN9XQ~
\end{lstlisting}

\begin{lstlisting}[caption={Decoded disclosures}, label={lst:disclosurelistdecoded}]
['avy3ozSTqLZDMLlUHQI8jQ', {'path': 'example\\test1.txt', 'offset': 0, 'length': 36, 'hash': 'hwKmdFRN69WkkKB9ZeFmDnbwc5yzxu9J0vxI2I-KQro', 'key': 'RvPUIIaLLZ2Oajh0f5EmxqzkTv1hffiTDLzeZmD3lNE', 'alg': 3}]

['lWhihnN-FDe0VtLGEm4a8g', {'path': 'example\\test2.txt', 'offset': 36, 'length': 36, 'hash': 'jPrDtbgHl5dyFYeRzCbrHjjR0WiSuGFKF5InAz-BI8g', 'key': 'JLhRPpAGl0tU3aSCcK0_Ep2-1OLlmeG3k2OxtbGXRsU', 'alg': 3}]

['yDaYXZNst8--At2q-nhkgQ', {'path': 'example\\subd\\test3.txt', 'offset': 72, 'length': 35, 'hash': 'AYxnosMO2Zd-Hd3-mMrFQhZdrDVc-XZMkaNiYT51KTM', 'key': 'hxuKT_pA0Pf9e1z6kjty-4zALJ5KdhtmNXlowVRIQ64', 'alg': 3}]
\end{lstlisting}

\subsection*{Verifier reconstruction}
The issuer creates the signed payload as represented in listing \ref{lst:signedjwtpayload} and the disclosure list as in listing \ref{lst:disclosurelist}. The disclosure digests in the payload are created by hashing the Base64 Encoded disclosures as shown in listing \ref{lst:disclosurelist}. The listings \ref{lst:signedjwtpayload}, \ref{lst:jwtpayload}, \ref{lst:disclosurelist}, and \ref{lst:disclosurelistdecoded} shows compliance with the SD-JWT standard as defined in \cite{rfc9901}.

The verifier extracts the JWS signature from the JWT payload and validates the same with the issuer public key. For each disclosure, the verifier computes BASE64URL(SHA-256(ASCII(Disclosure)), where Disclosure is the Base64Url encoded disclosure string from Listing \ref{lst:disclosurelist}. The resulting digest is compared with the corresponding digest placeholder in the issuer-signed JWT Payload. Once they are validated, the disclosures are Base64Url decoded to retrieve the metadata required to reconstruct the file, and the verifier inspects the offset and length fields of the same. The verifier then extracts the corresponding segment marked by the offset and length from the Blob. It then decrypts and extracts the nonce and AEAD tag from the section. It uses the AES key in disclosure to decrypt the encrypted file and validates the AEAD tag. A hash of the decrypted file is calculated and validated against the hash in the disclosure payload. It finally inspects the file name and file types from the path metadata field in the disclosure and writes the file to the disk for usage.

\section{Test Vectors}

This section highlights the test vectors from the sample implementation that can be used to reproduce the study.

\begin{lstlisting}[caption={Input files}, label={lst:inputfiles}]
File 1:
    Path: example\test1.txt
    Plaintext content: test1
    Plaintest SHA-256 hash (URL Safe B64 Encoded): hwKmdFRN69WkkKB9ZeFmDnbwc5yzxu9J0vxI2I-KQro
    
File 2:
    Path: example\test2.txt
    Plaintext content: test2
    Plaintest SHA-256 hash (URL Safe B64 Encoded): jPrDtbgHl5dyFYeRzCbrHjjR0WiSuGFKF5InAz-BI8g
    
File 3:
    Path: example\subd\test3.txt
    Plaintext content: test3
    Plaintest SHA-256 hash (URL Safe B64 Encoded): AYxnosMO2Zd-Hd3-mMrFQhZdrDVc-XZMkaNiYT51KTM
\end{lstlisting}

\begin{lstlisting}[caption={Issuer Parameters}, label={lst:issuerparams}]
Signing algorithm: ES256
Issuer pulic key (COSE Encoded):
    {
      "crv": "P-256",
      "kty": "EC",
      "x": "Ie9X6X9D2m8Oh3k1WXmyJNR_aZpROb7DDpDBstFc9yQ",
      "y": "QFQmYn0jObNveUyUbXnrIuKVfjbUac88J7P1AWc9yoM"
    }
Issuer private key (COSE Encoded):
    {
      "crv": "P-256",
      "d": "NyXgWlmQ-PzWi0dzOBWC168MTPXXt0R98Q3AAtnroQI",
      "kty": "EC",
      "x": "Ie9X6X9D2m8Oh3k1WXmyJNR_aZpROb7DDpDBstFc9yQ",
      "y": "QFQmYn0jObNveUyUbXnrIuKVfjbUac88J7P1AWc9yoM"
    }
_sd_alg value: sha256

\end{lstlisting}

\begin{lstlisting}[caption={Per file disclosure data}, label={lst:perfiledisclsoure}]
File 1:
    Salt: avy3ozSTqLZDMLlUHQI8jQ
    File descriptor before encoding: {'path': 'example\\test1.txt', 'offset': 0, 'length': 36, 'hash': 'hwKmdFRN69WkkKB9ZeFmDnbwc5yzxu9J0vxI2I-KQro', 'key': 'RvPUIIaLLZ2Oajh0f5EmxqzkTv1hffiTDLzeZmD3lNE', 'alg': 3}
    Base64URL encoded disclosure: WyJhdnkzb3pTVHFMWkRNTGxVSFFJOGpRIiwgeyJwYXRoIjogImV4YW1wbGVcXHRlc3QxLnR4dCIsICJvZmZzZXQiOiAwLCAibGVuZ3RoIjogMzYsICJoYXNoIjogImh3S21kRlJONjlXa2tLQjlaZUZtRG5id2M1eXp4dTlKMHZ4STJJLUtRcm8iLCAia2V5IjogIlJ2UFVJSWFMTFoyT2FqaDBmNUVteHF6a1R2MWhmZmlUREx6ZVptRDNsTkUiLCAiYWxnIjogM31d
    Disclosure digest: DXpSzubBkS1dHbcqaUT6oVUV8HVAmyYuwzOpeOM-mxo

File 2:
    Salt: lWhihnN-FDe0VtLGEm4a8g
    File descriptor before encoding: {'path': 'example\\test2.txt', 'offset': 36, 'length': 36, 'hash': 'jPrDtbgHl5dyFYeRzCbrHjjR0WiSuGFKF5InAz-BI8g', 'key': 'JLhRPpAGl0tU3aSCcK0_Ep2-1OLlmeG3k2OxtbGXRsU', 'alg': 3}
    Base64URL encoded disclosure: WyJsV2hpaG5OLUZEZTBWdExHRW00YThnIiwgeyJwYXRoIjogImV4YW1wbGVcXHRlc3QyLnR4dCIsICJvZmZzZXQiOiAzNiwgImxlbmd0aCI6IDM2LCAiaGFzaCI6ICJqUHJEdGJnSGw1ZHlGWWVSekNickhqalIwV2lTdUdGS0Y1SW5Bei1CSThnIiwgImtleSI6ICJKTGhSUHBBR2wwdFUzYVNDY0swX0VwMi0xT0xsbWVHM2syT3h0YkdYUnNVIiwgImFsZyI6IDN9XQ
    Disclosure digest: EEZmZLkr1zN576F949D6FpHrw6cnOCN0Spq1v56eYmM
    
File 3:
    Salt: yDaYXZNst8--At2q-nhkgQ
    File descriptor before encoding: {'path': 'example\\subd\\test3.txt', 'offset': 72, 'length': 35, 'hash': 'AYxnosMO2Zd-Hd3-mMrFQhZdrDVc-XZMkaNiYT51KTM', 'key': 'hxuKT_pA0Pf9e1z6kjty-4zALJ5KdhtmNXlowVRIQ64', 'alg': 3}
    Base64URL encoded disclosure: WyJ5RGFZWFpOc3Q4LS1BdDJxLW5oa2dRIiwgeyJwYXRoIjogImV4YW1wbGVcXHN1YmRcXHRlc3QzLnR4dCIsICJvZmZzZXQiOiA3MiwgImxlbmd0aCI6IDM1LCAiaGFzaCI6ICJBWXhub3NNTzJaZC1IZDMtbU1yRlFoWmRyRFZjLVhaTWthTmlZVDUxS1RNIiwgImtleSI6ICJoeHVLVF9wQTBQZjllMXo2a2p0eS00ekFMSjVLZGh0bU5YbG93VlJJUTY0IiwgImFsZyI6IDN9XQ
    Disclosure digest: 8Q-8ucovy_NPfs_VIVc492X2aeWD-NSWms6dWwbq_UU
\end{lstlisting}

\begin{lstlisting}[caption={Encryption data}, label={lst:encryptiondata}]
File 1:
    AES Key: RvPUIIaLLZ2Oajh0f5EmxqzkTv1hffiTDLzeZmD3lNE
    Nonce: MGt5DBiJJPA5f597
    Ciphertext: XVC1ZDvaBwA
    AEAD tag: zpAtk7ko7Zfmgz1NiAoCBA
    Offset: 0
    Length: 36
    
File 2:
    AES Key: JLhRPpAGl0tU3aSCcK0_Ep2-1OLlmeG3k2OxtbGXRsU
    Nonce: 9f-FrWVGtXYFBrMy
    Ciphertext: NXVyIFQc75A
    AEAD tag: wcwibvwmhwqE9ghr5fNyVQ
    Offset: 36
    Length: 36
    
File 3:
    AES Key: AYxnosMO2Zd-Hd3-mMrFQhZdrDVc-XZMkaNiYT51KTM
    Nonce: ac3gjpJT76kg_eEu
    Ciphertext: BLC5O2W_Sw
    AEAD tag: nk9brTAt5L1nird-kDbY0Q
    Offset: 72
    Length: 35
    
Blob: data:application/octet-stream;base64,XVC1ZDvaBwAwa3kMGIkk8Dl_n3vOkC2TuSjtl-aDPU2ICgIENXVyIFQc75D1_4WtZUa1dgUGszLBzCJu_CaHCoT2CGvl83JVBLC5O2W_S2nN4I6SU--pIP3hLp5PW60wLeS9Z4q3fpA22NE=
\end{lstlisting}

\subsection*{Expected outputs}
\begin{lstlisting}[caption={Decoded JWT Payload}, label={lst:decodedjwtpayload}]
{
  "files": [
    {
      "...": "DXpSzubBkS1dHbcqaUT6oVUV8HVAmyYuwzOpeOM-mxo"
    },
    {
      "...": "EEZmZLkr1zN576F949D6FpHrw6cnOCN0Spq1v56eYmM"
    },
    {
      "...": "8Q-8ucovy_NPfs_VIVc492X2aeWD-NSWms6dWwbq_UU"
    }
  ],
  "blob": "data:application/octet-stream;base64,XVC1ZDvaBwAwa3kMGIkk8Dl_n3vOkC2TuSjtl-aDPU2ICgIENXVyIFQc75D1_4WtZUa1dgUGszLBzCJu_CaHCoT2CGvl83JVBLC5O2W_S2nN4I6SU--pIP3hLp5PW60wLeS9Z4q3fpA22NE=",
  "_sd_alg": "sha-256"
}
\end{lstlisting}

\begin{lstlisting}[caption={Issuer signed JWT and Final presentation}, label={lst:signedjwtpresentation}]
Compact issuer-signed JWT: eyJhbGciOiAiRVMyNTYiLCAidHlwIjogImV4YW1wbGUrc2Qtand0In0.eyJmaWxlcyI6IFt7Ii4uLiI6ICJEWHBTenViQmtTMWRIYmNxYVVUNm9WVVY4SFZBbXlZdXd6T3BlT00tbXhvIn0sIHsiLi4uIjogIkVFWm1aTGtyMXpONTc2Rjk0OUQ2RnBIcnc2Y25PQ04wU3BxMXY1NmVZbU0ifSwgeyIuLi4iOiAiOFEtOHVjb3Z5X05QZnNfVklWYzQ5MlgyYWVXRC1OU1dtczZkV3dicV9VVSJ9XSwgImJsb2IiOiAiZGF0YTphcHBsaWNhdGlvbi9vY3RldC1zdHJlYW07YmFzZTY0LFhWQzFaRHZhQndBd2Eza01HSWtrOERsX24zdk9rQzJUdVNqdGwtYURQVTJJQ2dJRU5YVnlJRlFjNzVEMV80V3RaVWExZGdVR3N6TEJ6Q0p1X0NhSENvVDJDR3ZsODNKVkJMQzVPMldfUzJuTjRJNlNVLS1wSVAzaExwNVBXNjB3TGVTOVo0cTNmcEEyMk5FPSIsICJfc2RfYWxnIjogInNoYS0yNTYifQ.4tSNrDAPBPZWYWurQw5k0pB1M5RXvXZukrkqtWYKf_Y1HigkbIbPNEqynordiV4Bqk44btZDFIf9GV_gDFxYOA

Final SD-JWT presentation: eyJhbGciOiAiRVMyNTYiLCAidHlwIjogImV4YW1wbGUrc2Qtand0In0.eyJmaWxlcyI6IFt7Ii4uLiI6ICJEWHBTenViQmtTMWRIYmNxYVVUNm9WVVY4SFZBbXlZdXd6T3BlT00tbXhvIn0sIHsiLi4uIjogIkVFWm1aTGtyMXpONTc2Rjk0OUQ2RnBIcnc2Y25PQ04wU3BxMXY1NmVZbU0ifSwgeyIuLi4iOiAiOFEtOHVjb3Z5X05QZnNfVklWYzQ5MlgyYWVXRC1OU1dtczZkV3dicV9VVSJ9XSwgImJsb2IiOiAiZGF0YTphcHBsaWNhdGlvbi9vY3RldC1zdHJlYW07YmFzZTY0LFhWQzFaRHZhQndBd2Eza01HSWtrOERsX24zdk9rQzJUdVNqdGwtYURQVTJJQ2dJRU5YVnlJRlFjNzVEMV80V3RaVWExZGdVR3N6TEJ6Q0p1X0NhSENvVDJDR3ZsODNKVkJMQzVPMldfUzJuTjRJNlNVLS1wSVAzaExwNVBXNjB3TGVTOVo0cTNmcEEyMk5FPSIsICJfc2RfYWxnIjogInNoYS0yNTYifQ.4tSNrDAPBPZWYWurQw5k0pB1M5RXvXZukrkqtWYKf_Y1HigkbIbPNEqynordiV4Bqk44btZDFIf9GV_gDFxYOA~WyJhdnkzb3pTVHFMWkRNTGxVSFFJOGpRIiwgeyJwYXRoIjogImV4YW1wbGVcXHRlc3QxLnR4dCIsICJvZmZzZXQiOiAwLCAibGVuZ3RoIjogMzYsICJoYXNoIjogImh3S21kRlJONjlXa2tLQjlaZUZtRG5id2M1eXp4dTlKMHZ4STJJLUtRcm8iLCAia2V5IjogIlJ2UFVJSWFMTFoyT2FqaDBmNUVteHF6a1R2MWhmZmlUREx6ZVptRDNsTkUiLCAiYWxnIjogM31d~WyJsV2hpaG5OLUZEZTBWdExHRW00YThnIiwgeyJwYXRoIjogImV4YW1wbGVcXHRlc3QyLnR4dCIsICJvZmZzZXQiOiAzNiwgImxlbmd0aCI6IDM2LCAiaGFzaCI6ICJqUHJEdGJnSGw1ZHlGWWVSekNickhqalIwV2lTdUdGS0Y1SW5Bei1CSThnIiwgImtleSI6ICJKTGhSUHBBR2wwdFUzYVNDY0swX0VwMi0xT0xsbWVHM2syT3h0YkdYUnNVIiwgImFsZyI6IDN9XQ~WyJ5RGFZWFpOc3Q4LS1BdDJxLW5oa2dRIiwgeyJwYXRoIjogImV4YW1wbGVcXHN1YmRcXHRlc3QzLnR4dCIsICJvZmZzZXQiOiA3MiwgImxlbmd0aCI6IDM1LCAiaGFzaCI6ICJBWXhub3NNTzJaZC1IZDMtbU1yRlFoWmRyRFZjLVhaTWthTmlZVDUxS1RNIiwgImtleSI6ICJoeHVLVF9wQTBQZjllMXo2a2p0eS00ekFMSjVLZGh0bU5YbG93VlJJUTY0IiwgImFsZyI6IDN9XQ~

Selected Disclosures: ~WyJhdnkzb3pTVHFMWkRNTGxVSFFJOGpRIiwgeyJwYXRoIjogImV4YW1wbGVcXHRlc3QxLnR4dCIsICJvZmZzZXQiOiAwLCAibGVuZ3RoIjogMzYsICJoYXNoIjogImh3S21kRlJONjlXa2tLQjlaZUZtRG5id2M1eXp4dTlKMHZ4STJJLUtRcm8iLCAia2V5IjogIlJ2UFVJSWFMTFoyT2FqaDBmNUVteHF6a1R2MWhmZmlUREx6ZVptRDNsTkUiLCAiYWxnIjogM31d~WyJsV2hpaG5OLUZEZTBWdExHRW00YThnIiwgeyJwYXRoIjogImV4YW1wbGVcXHRlc3QyLnR4dCIsICJvZmZzZXQiOiAzNiwgImxlbmd0aCI6IDM2LCAiaGFzaCI6ICJqUHJEdGJnSGw1ZHlGWWVSekNickhqalIwV2lTdUdGS0Y1SW5Bei1CSThnIiwgImtleSI6ICJKTGhSUHBBR2wwdFUzYVNDY0swX0VwMi0xT0xsbWVHM2syT3h0YkdYUnNVIiwgImFsZyI6IDN9XQ~WyJ5RGFZWFpOc3Q4LS1BdDJxLW5oa2dRIiwgeyJwYXRoIjogImV4YW1wbGVcXHN1YmRcXHRlc3QzLnR4dCIsICJvZmZzZXQiOiA3MiwgImxlbmd0aCI6IDM1LCAiaGFzaCI6ICJBWXhub3NNTzJaZC1IZDMtbU1yRlFoWmRyRFZjLVhaTWthTmlZVDUxS1RNIiwgImtleSI6ICJoeHVLVF9wQTBQZjllMXo2a2p0eS00ekFMSjVLZGh0bU5YbG93VlJJUTY0IiwgImFsZyI6IDN9XQ~
\end{lstlisting}

\begin{lstlisting}[caption={Reconstructed file hashes}, label={lst:filehashes}]
File 1: hwKmdFRN69WkkKB9ZeFmDnbwc5yzxu9J0vxI2I-KQro
File 2: jPrDtbgHl5dyFYeRzCbrHjjR0WiSuGFKF5InAz-BI8g
File 3: AYxnosMO2Zd-Hd3-mMrFQhZdrDVc-XZMkaNiYT51KTM
\end{lstlisting}

The listings \ref{lst:inputfiles} to \ref{lst:filehashes} represent the test vectors which can be used to reproduce the study.

\section{Conclusion}
The proposed Selective File Disclosure System aims at solving the problem of verifiable file disclosure in a privacy-focused manner. It ensures only the authorized holders of the presentations and disclosures can view the contents of respective files without involving complex architectures like authentication systems, user databases, Identity and Access Management, etc. A few notable use cases of SFDS might be:

\begin{itemize}
    \item \textbf{Education.} Students often need to share verifiable copies of their transcripts, exam scripts and so on. Current SD-JWT standards may be used to share verifiable copies of the transcript where the data are represented in a machine-readable format, but not the entire scanned copies of the transcript or exam scripts. SFDS is a promising solution to the problem.
    \item \textbf{Medicine.} Like the problem of education, medicine demands the authenticity and integrity of health information. Current SD-JWT standards offer sharing medical reports, containing medical conditions, values of various tests, and diagnoses in machine-readable formats. However, it does not support sharing raw reports like X-ray plates, images captured by MRI or Ultrasonography machines, etc. Further, these machines are often legacy and do not support digital signatures on the produced files. The SFDS aims to solve these and related problems.
\end{itemize}

\section*{Acknowledgment}

This research was conducted within the framework of the Erasmus Mundus Joint Master’s Programme in Applied Cybersecurity (CyberMACS) Project No. 101082683, co-funded by the European Union under the Erasmus+ MUNDUS Programme.

\bibliographystyle{IEEEtran}
\bibliography{ref}

\end{document}